# The optical phonoelectric effect

D. Choi[1], M. Först[1], M. Fechner[1], M. Buzzi[1], X. Deng[1,2], Z. Zeng[1,2], K. H. Martens[1], D. Prabhakaran[2], C. Putzke[1], P. Moll[1], P.G. Radaelli[2], A. Cavalleri[1,2]

*[1]Max Planck Institute for the Structure and Dynamics of Matter, Hamburg, Germany*
*[2]Department of Physics, Clarendon Laboratory, University of Oxford, Oxford, United Kingdom*

**Piezoelectricity is a technologically important property of certain insulators in which mechanical strain induces an electrical polarization. However, the rate at which a piezoelectric response can be established over a macroscopic volume is limited by the sound velocity, constraining applications in high-bit-rate transduction and sensing. Furthermore, the strength of the piezoelectric effect is not readily tunable, as it depends on intrinsic anharmonic coupling between strain and intra-unit-cell distortions in a given material. Lastly, the maximum amplitude of the effect is bounded by material fracture, which sets in already at percent level strain values. Here we overcome these limitations by realizing a strain-free, piezoelectric-like response driven solely by photo-excited optical phonon distortions. We demonstrate such optical phonoelectricity in the weak piezoelectric $BPO_4$, in which we induce electrical polarization through phonon rectification. This effect is established over macroscopic volumes with four orders of magnitude higher speed than piezoelectric responses, ultimately limited by the speed of light. The maximum induced polarization is estimated to be far in excess of that attainable through strain at the fracture limit. Ultrafast phonoelectricity opens up new opportunities for optical control in quantum materials, but also for device applications.**

Piezoelectricity is based on the linear coupling between mechanical stress and electrical polarization, with applications spanning actuators, sensors, and transducers[1–4]. Despite its broad utility, piezoelectricity is subject to several fundamental limitations, especially in speed and in the possibility to engineer the transduction strength.

Here, we explore an ultrafast and strain-free route to induce electrical polarization in a piezoelectric material. We focus on Boron phosphate (BPO$_4$, space group I$\bar{4}$), a non-polar crystal composed of interpenetrating layers of BO$_4$ and PO$_4$ tetrahedra stacked along the crystal *c* axis (Fig. 1a, left panel). In equilibrium, dipoles associated with the P–O and B–O bonds compensate each other, yielding zero net macroscopic polarization (Fig. 1a, right panel). When uniaxial stress is applied perpendicular to the stacking direction, anisotropic lattice distortions induce uncompensated electric dipoles on the two tetrahedral substructures (Fig. 1b), resulting in a finite net polarization. Reversing the applied stress produces electrical polarization of opposite sign. However, because of the weak coupling between applied stress and the lattice distortions responsible for polarization, the piezoelectric effect of BPO$_4$ is modest (0.13 µC cm$^{-2}$ GPa$^{-1}$), and smaller than that of other piezoelectrics such as GaN[5,6] (~0.4 µC cm$^{-2}$ GPa$^{-1}$), AlN[6] (~0.5 µC cm$^{-2}$ GPa$^{-1}$), and ZnO[7] (~1.2 µC cm$^{-2}$ GPa$^{-1}$).

Second-harmonic generation (SHG) was used to validate the existence of the phonoelectric effect. First, we evaluated the static SHG polarimetry in equilibrium non-polar BPO$_4$, which was measured on a (010)-surface cut crystal by rotating the polarization of incident near-infrared probe pulses in the *ac* plane and by detecting the generated SHG intensity for fixed polarizations along the *a* and *c* axes. The results show excellent agreement with calculations performed in the parent point group $\bar{4}$ (see Supplementary Information). Crucially, the $\chi_{33}$ tensor element vanishes in the non-polar structure.

Resonantly driven infrared-active phonons couple anharmonically to one or more vibrational modes, resulting in a displaced crystal structure. In the past, nonlinear phononics has proven effective in controlling a broad range of ferroic orders, including ferroelectricity[8,9], piezomagnetism[10], ferromagnetism[10,11], ferroaxiality[12], and chirality[13]. For piezoelectrics, nonlinear phononic control is attractive as it redistributes intra-unit-cell ionic positions without altering the lattice constants, and is thus not constrained by the speed of sound or by mechanical fracture limits. Furthermore, unlike piezoelectricity, by tuning the wavelength or polarization of the excitation, one can select different phonon resonances and engineer the strength of the process.

To induce a phonoelectric polarization in $BPO_4$, an infrared-active phonon of E-symmetry and amplitude $Q_E$, polarized along the *a* or *b* axis is resonantly driven by a mid-infrared pulse. It couples anharmonically to B-symmetry modes with amplitude $Q_B$, polarized along the *c* axis, through an interaction potential term $V_{nonl.} = \beta Q_E^2 Q_B$. The coupling leads to rectification of the B-symmetry modes, reducing the crystal symmetry from point group $\bar{4}$ to 2 and inducing an electrical polarization along the *c* axis [14]. Figure 1c illustrates how this mechanism generates a finite polarization by distorting the $BO_4$ and $PO_4$ tetrahedra, contrast to the piezoelectric response in Fig. 1b. Polarization in opposite direction along the *c* axis is achieved by resonantly driving the doubly degenerate E mode along the other in-plane axis, as also the piezoelectric responses along the a- and b-axes are opposite in sign. Note that although the piezoelectric and the optical phonoelectric effects both involve B-mode displacements, the signs and relative magnitudes of the coupling coefficients differ markedly between the two mechanisms. In the piezoelectric effect, the polarization resulting from the superposition of four B-mode displacements nearly cancel each other, yielding an almost vanishing net polarization. By contrast, the phonoelectric effect provides a different combination of B-mode displacements, producing a substantial net electrical polarization (see Supplementary Information).

In the time-resolved SHG experiments, the E-mode along the *a* axis was resonantly excited with mid-infrared pulses of up to 14.2 mJ cm$^{-2}$ fluence at 28 THz center frequency. The polarization of the 800-nm wavelength probe pulses was rotated in the *ac* plane, and 400-nm SHG polarized along the *c* axis was detected to evidence the light-induced polar structure (Fig. 2a).

The SHG intensity $I_c \propto \chi_{33}^2 E_{probe,c}^4$ is sensitive to electrical polarization along the *c* axis through the nonlinear tensor element $\chi_{33}$, which is forbidden in the parent point group $\bar{4}$ and becomes finite in the lower-symmetry point group 2. As shown in Figs. 2b and 2c, the optical excitation induced a strong SHG intensity when both the probe and SHG polarizations are aligned along the crystal *c* axis, reflecting an induced $\chi_{33}$ tensor element. At the probe polarization of $\theta = 0°$, where $\theta$ denotes the angle measured from the crystal *c* axis, the rectified SHG signal $I_{rect}^{(2\omega)}(\theta = 0°)$ is proportional to $(\delta\chi_{33})^2$, with $\delta\chi_{33}$ representing the finite light-induced $\chi_{33}$ tensor element. Because $\delta\chi_{33}$ scales with $Q_B \propto Q_E^2 \propto E_{pump}^2$, one expects $I_{rect}^{(2\omega)}(\theta = 0°) \propto E_{pump}^4 \propto F^2$ where $F$ denotes the excitation fluence.

SHG polarimetry at the peak of the transient response was acquired for three fluences of the mid-infrared excitation pulses between 3.1 and 14.2 mJ cm$^{-2}$. Figure 3a shows the corresponding probe-polarization-dependent amplitudes of the rectified SHG intensity changes (see Supplementary Information). As the excitation fluence increases, the dipolar SHG contribution along the crystal $c$ axis grows faster than the fourfold-symmetric component along the diagonals.

At the probe polarization of 45°, the signal is dominated by the homodyne hyper-Raman response of the resonantly driven mode $Q_E$, and the heterodyne mixing of SHG fields generated by the light-induced $\delta\chi_{33}$ with SHG fields generated by the equilibrium $\chi_{31}$ tensor elements. The hyper-Raman response arises from inelastic scattering involving two probe photons and the driven phonon mode $Q_E$[15,16], which produces a nonlinear polarization $P_i(2\omega_{pr} \pm \Omega) = \frac{\partial \chi_{ijk}^{(2)}}{\partial Q_E} Q_E(\Omega) E_j(\omega_{pr}) E_k(\omega_{pr})$, where $\Omega$ denotes the mode frequency and $E(\omega_{pr})$ the probe electric field. This effect induces a transient SHG intensity, which is not associated with the phonoelectric effect. Both hyper-Raman and heterodyne mixing scale as $E_{pump}^2$, leading to $I_{rect}^{(2\omega)}(\theta = 45°) \propto E_{pump}^2 \propto F$ (see Supplementary Information). The excitation fluence dependencies at $\theta = 0°$ and $\theta = 45°$ were experimentally verified in Fig. 3b. The polarization dependence of the discussed contributions over the full 360° rotation of $\theta$ accounts for the distinct growth behavior of the two- and four-fold pattern observed in the differential SHG polarimetry (Fig. 3a). Notable in these results is the speed of the effect. As mentioned above, because the optical excitations that induce lattice rearrangements propagate at the light speed of phonon polaritons (~40 μm ps$^{-1}$), macroscopic regions of the material can be polarized about four orders of magnitude faster than with strain-based approaches, which are limited by acoustic phonon propagation (~7 nm ps$^{-1}$).

Whilst the piezoelectric effect could not be observed by SHG, detection of a large photo-induced signal is already an indication of the enhanced strength of the optical phonoelectric effect. However, the absolute magnitude of the photo-induced electrical polarization could not be extracted from the measured SHG, owing to the lack of an independent calibration of $\chi_{33}$, which is absent in the equilibrium state and difficult to detect under an applied static strain. Instead, the induced electrical polarization was estimated by density functional theory (DFT) calculations, which yielded a value of 0.72 µC cm$^{-2}$ at the highest incident excitation fluence of 14.2 mJ cm$^{-2}$ used in our experiments (see Supplementary Information). Figure 4 shows that mechanical stress of approximately 5.4 GPa is required to generate an equivalent polarization via the piezoelectric effect, also estimated from DFT. Assuming a maximum elastic strain of 1%, this stress value significantly exceeds the crystal fracture limit of 1.2 GPa. Because the phonoelectric effect develops at constant volume, its magnitude is constrained only by the lattice stability under E mode excitation, likely determined by the Lindemann criterion (~10% of the lattice constant). In this limit, we estimate that a maximum phonoelectric polarization as high as 74 µC cm$^{-2}$ could be reached (see Supplementary Information), which is comparable to the spontaneous electrical polarization of representative ferroelectric materials such as BaTiO$_3$ (~26 µC cm$^{-2}$)[17] and BiFeO$_3$ (~146 µC cm$^{-2}$)[18,19].

The optical phono-electric effect offers an advantage in conversion efficiency – defined here as the energy density required to induce a given polarization amplitude – relative to the piezoelectric response. Because the required energy density scales linearly with the induced polarization for the phono-electric effect, but quadratically for the piezoelectric effect, the relative efficiency advantage of the phonoelectric mechanism becomes increasingly pronounced at larger polarization amplitudes. We define $r = w_{phono}/w_{piezo} = (F/\delta)/(1/2u\sigma)$ as the ratio between required energy from phonoelectric and piezoelectric effects. Here, $w$ denotes the input energy density, $F$ the excitation fluence, $\delta$ the extinction depth of the mid-IR excitation pulses, $\sigma$ the equivalent stress, and $u$ the resulting strain. In the Lindemann limit, $r$ is estimated to be about 0.01, suggesting that in BPO$_4$ phonoelectricity is two orders of magnitude more efficient than piezoelectricity. The use of narrowband excitation pulses would enhance the mode selectivity[20,21] and improve the overall conversion efficiency.

In summary, we demonstrated a photo-induced electrical polarization in BPO$_4$. The macroscopic volume of the crystal is polarized with a speed approximately four orders of magnitude faster than in the piezoelectric response. Furthermore, DFT estimates a polarization of 0.72 μC cm$^{-2}$ at the highest incident excitation fluence of 14.2 mJ cm$^{-2}$, exceeding that attainable under stress of 1.2 GPa at the fracture-limited maximum elastic strain of 1%. This enhancement arises, because the phonoelectric effect provides a strong coupling to lattice distortions and is not constrained by mechanical fracture limit. The phono-electric effect may provide a platform for applications, in which electromagnetic signals are converted into electrical signal, such as sensors or transducers. More broadly, this approach may yield new opportunities for achieving enhanced electrical responses and high-speed operations in microelectronics, memory devices, sensors, and quantum computation systems.

**Figures**

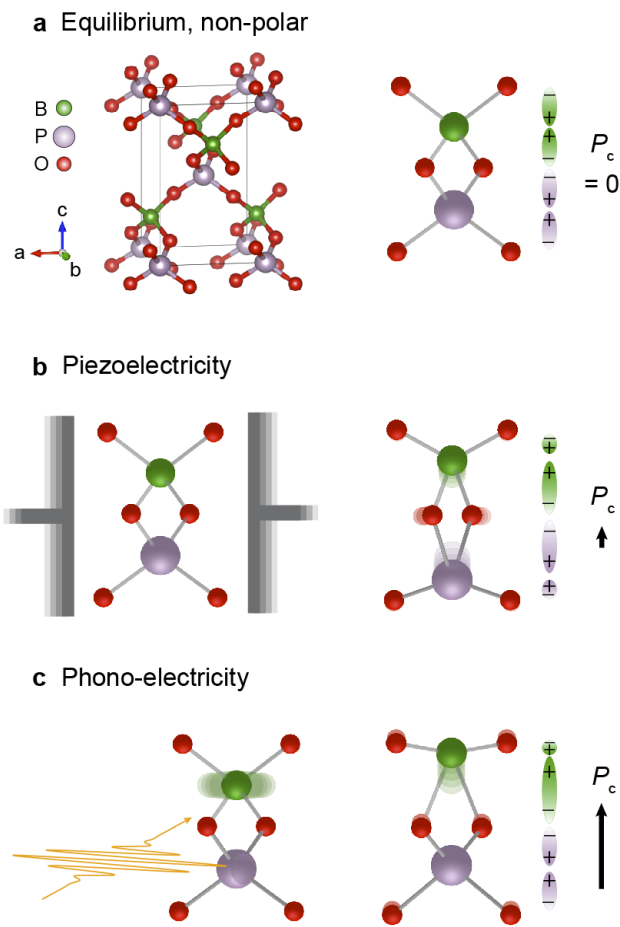

**Fig. 1. Electrical polarizations induced by the piezoelectric and phono-electric effects. a**, Crystal lattice of BPO$_4$ (left) and a sub-unit-cell structure (right). Dipoles on the BO$_4$ and PO$_4$ tetrahedra cancel, resulting in zero net polarization. **b,** Symmetry breaking induced by the piezoelectric effect. Compressive stress along the *a* axis produces an anisotropic lattice distortion, which generates an imbalance of *c*-axis dipoles with opposite directions within the PO$_4$ and BO$_4$ units. Because P–O bonds contribute a larger polarization than B–O bonds, a small, but finite net polarization emerges. **c**, Symmetry breaking induced by the phonoelectric effect. Resonant optical excitation of an E-symmetry mode along the *a* axis rectifies *c*-axis polarized B-symmetry modes through anharmonic coupling, resulting in a finite polarization. The polarization generated within the BO$_4$ units dominates over that from the PO$_4$ units, yielding a large net polarization.

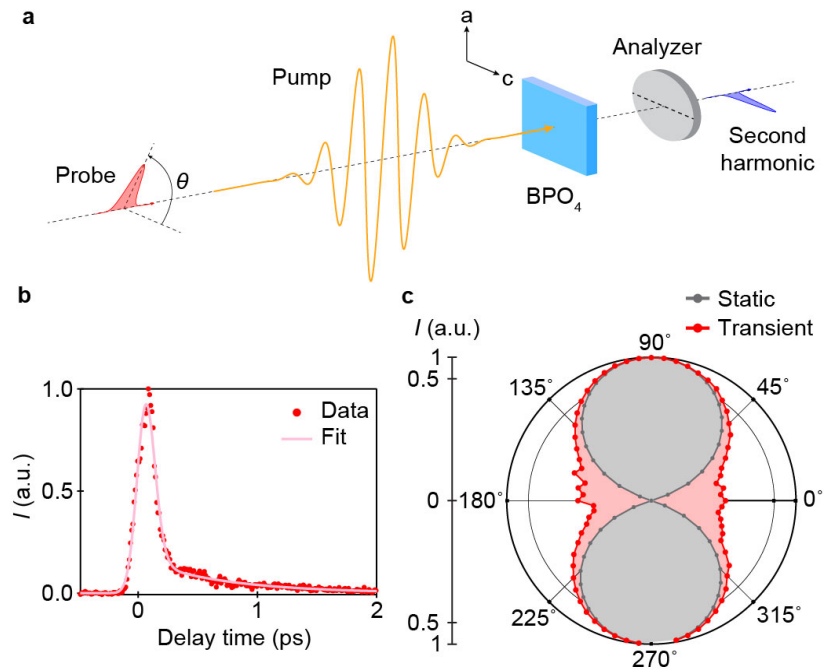

**Fig. 2. Photo-induced electrical polarization probed by second-harmonic generation. a**, Experimental configuration. A linearly polarized mid-infrared pulse, polarized along the BPO$_4$ *a* axis, induces a finite electrical polarization that is detected via second-harmonic generation of time-delayed near-infrared pulse probes. The experiment was carried out as a function of polarization of the incident probe, and the SHG intensity was collected with its polarization parallel to the *c* axis. **b**, Photo-induced SHG intensity (dots) with *c*-axis polarized probe and SHG light. The solid line is a fit to the data. The strong, rapidly decaying component is attributed to the transient electrical polarization induced by the phonoelectric effect, whilst the slowly decaying component results from the resonantly driven phonon. **c**, Probe polarization dependent SHG intensity at the time delay of the maximum photo-induced response (red), together with the equilibrium SHG polarimetry (grey). A strong light-induced enhancement is observed when the probe polarization is aligned close to the crystal *c* axis, evidencing a finite electrical polarization along this direction.

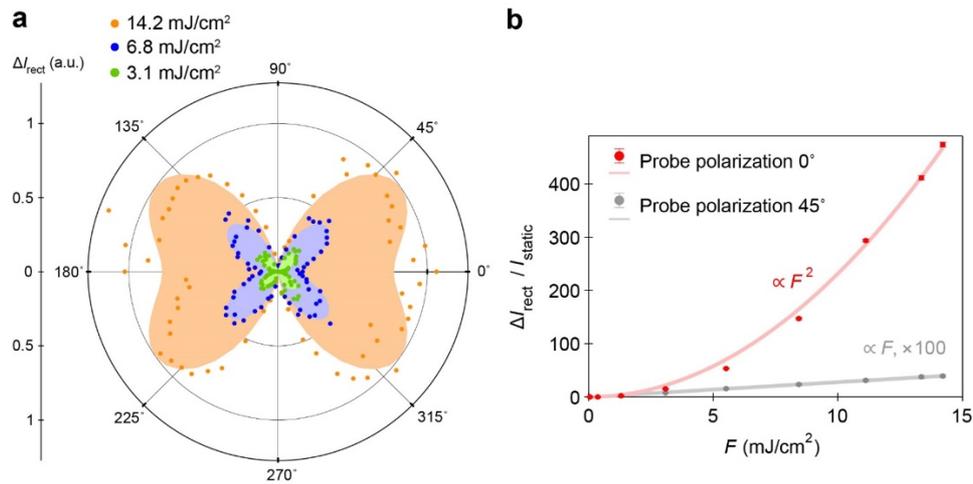

**Fig. 3. Fluence dependent characterization of the photo-induced polar state. a**, Polarimetry of the rectified component of the photo-induced changes in SHG intensity, extracted from fits to the time-resolved experimental data at each polarization angle, for three excitation fluences **b,** Excitation fluence dependence of the rectified component of the photo-induced SHG intensity changes measured at probe polarizations of 0° (red) and 45° (gray) with respect to the crystal *c* axis. Data points at the probe polarization of 0° are divided by the standard error of the static SHG intensity, since the static SHG intensity vanishes.

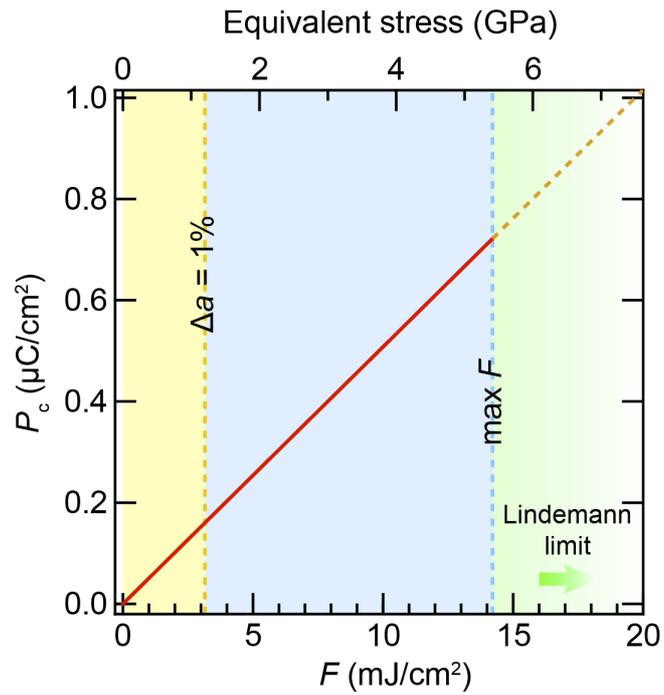

**Fig. 4. Comparison between the phonoelectric and piezoelectric effects.** The electrical polarization $P_c$ along the $c$ axis, estimated by density functional theory (see Supplementary Information). At the highest excitation fluence of 14.2 mJ cm$^{-2}$ (dashed blue line), it reaches a maximum value of 0.72 µC cm$^{-2}$. Achieving an equivalent polarization through the piezoelectric effect would require a mechanical stress of approximately 5.4 GPa. This value exceeds the crystal fracture limit (~1.2 GPa, yellow dashed line), assuming a maximum elastic strain of 1%. Furthermore, the Lindemann stability limit for the driven intra-unit cell distortions (corresponding to an excitation fluence of ~1.5 J cm$^{-2}$, outside the range of this plot) sets an upper bound on the polarization achievable via the phonoelectric effect, yielding estimated values of 74 µC cm$^{-2}$. For comparison, the spontaneous electrical polarization of representative ferroelectric materials such as BaTiO$_3$ (~26 µC cm$^{-2}$)[17] and BiFeO$_3$ (~146 µC cm$^{-2}$)[18,19].

**Acknowledgement**

We thank E. Amuah for assistance in the optical experiments. The work received funding from the Cluster of Excellence ´CUI: Advanced Imaging of Matter´ of the Deutsche Forschungsgemeinschaft (DFG), EXC 2056, project ID 390715994. D.C. was supported by a fellowship from the Alexander von Humboldt Foundation.

# Supplementary Information for

## The optical phonoelectric effect

D. Choi[1], M. Först[1], M. Fechner[1], M. Buzzi[1], X. Deng[1,2], Z. Zeng[1,2], K. H. Martens[1], D. Prabhakaran[2], C. Putzke[1], P. Moll[1], P.G. Radaelli[2], A. Cavalleri[1,2]

[1]*Max Planck Institute for the Structure and Dynamics of Matter, Hamburg, Germany*
[2]*Department of Physics, Clarendon Laboratory, University of Oxford, Oxford, United Kingdom*

## S1. Methods and sample preparation

### S1.1. Experimental setup

The experimental setup, sketched in Fig. S1.1, was powered by a Ti:Sapphire amplifier system delivering 35-fs pulses, centered at 800 nm wavelength, at a repetition rate of 1 kHz.

It was used to drive two dual-stage optical parametric amplifiers (OPAs), which were seeded by the same white light. The signal outputs at 1.37 µm and 1.21 µm were mixed in a nonlinear GaSe crystal to generate mid-infrared pulses centered at 28.4 THz with a full-width at half-maximum of 3.3 THz (see Fourier transform infrared spectroscopy characterization in Fig. S1.2). They were focused onto the sample using an off-axis parabolic mirror with a focal length of 50 mm. The maximum excitation fluence available in the experiment was 14.2 mJ/cm$^2$.

The 800-nm wavelength probe pulses were aligned collinear with the mid-infrared excitation. Their polarization was rotated in the sample surface by the combination of a quarter-wave plate and a polarizer. The generated second-harmonic radiation, centered at 400 nm, was isolated from the fundamental light by a dichroic mirror and two bandpass filters. A second polarizer selected its polarization state before detection in a photomultiplier tube.

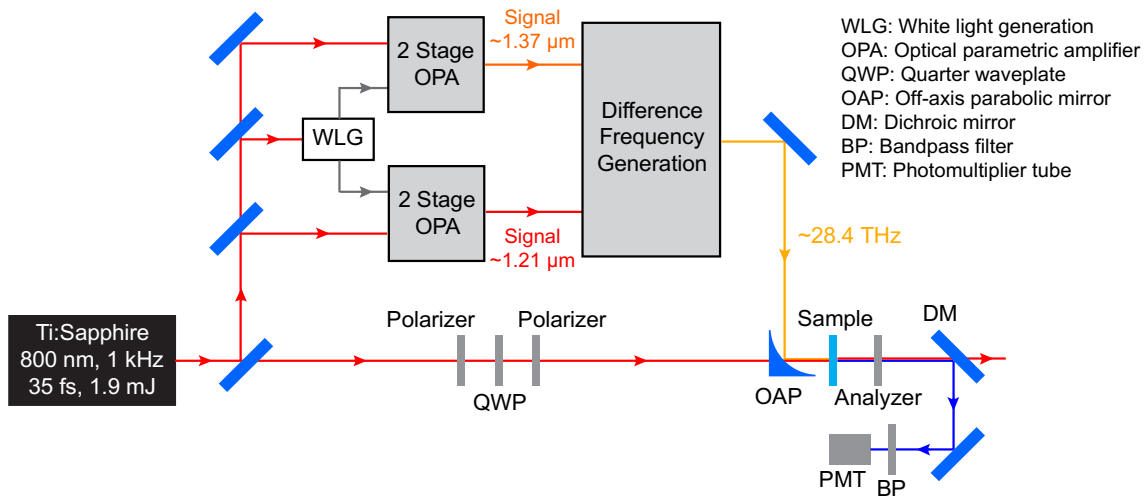

**Fig. S2.1. Sketch of the experimental setup.**

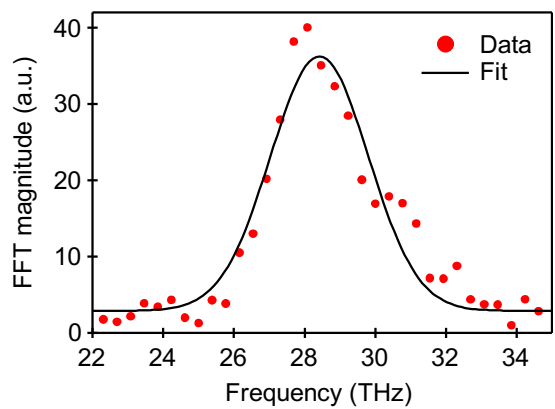

**Fig. S1.2. Spectrum of the excitation pulses characterized by Fourier-transform infrared spectroscopy.**

### S1.2. Sample growth and preparation

The BPO$_4$ single crystal was synthesized and prepared for the optical experiments as described in the Supplementary Materials of Reference 1, however an optically flat (010) surface was chosen for the experiments presented in this manuscript. The sample thickness was approximately 130 µm.

## S2. Second Harmonic Generation

### S2.1. Equilibrium Second harmonic generation polarimetry

The second-order nonlinear polarization $P_i$ in a solid is given by $P_i(2\omega) = \chi_{ijk}(2\omega;\omega,\omega)E_j(\omega)E_k(\omega)$, with the third-rank susceptibility tensor $\chi(2\omega;\omega,\omega)_{ijk}$ and the electric field $E(\omega)$ at the fundamental frequency $\omega$. In the commonly used Voight notation, the second harmonic generation (SHG) susceptibility tensors of the BPO$_4$ equilibrium point group $\bar{4}$ reads:

$$\text{Point group } \bar{4}: \quad \chi(2\omega;\omega,\omega)_{\alpha\beta} = \begin{bmatrix} 0 & 0 & 0 & \chi_{14} & \chi_{15} & 0 \\ 0 & 0 & 0 & -\chi_{15} & \chi_{14} & 0 \\ \chi_{31} & -\chi_{31} & 0 & 0 & 0 & \chi_{36} \end{bmatrix}, \quad (S2.1)$$

For a probe polarization oriented in the *ac* sample surface at an angle $\theta$ with respect to the *c* axis, the second-harmonic nonlinear polarization component along the crystal *c* axis is $P_c(2\omega) \propto \chi_{31} E_0^2(\omega) \sin^2\theta$, where $E_0(\omega)$ denotes the probe electric field amplitude. The detected SHG intensity with polarization parallel to the c axis follows

$$I_c \propto \chi_{31}^2 \sin^4\theta. \quad (S2.2)$$

For SHG detection along the *a* axis, the nonlinear polarization is $P_a(2\omega) \propto \chi_{15} E_0^2(\omega) \sin\theta \cos\theta$, hence

$$I_a \propto \chi_{15}^2 \sin^2\theta \cos^2\theta. \quad (S2.3)$$

Figure S2.1 presents experimental data, taken on the ac-surface plane of our BPO$_4$ sample in equilibrium, together with fits based on these expressions.

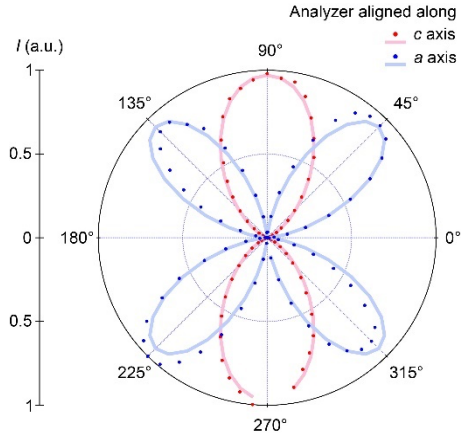

**Fig. S2.1. Static SHG polarimetry measurements.** Red and blue dots represent the experimental data. Solid lines are fits to the data using equations S2.2 and S2.3, respectively.

### S2.2. Time-resolved SHG polarimetry

To model the transient SHG polarimetry shown in Fig. 3a of the main text, we considered the SHG susceptibility of the photoinduced state as a perturbative expansion from point group $\bar{4}$ to point group 2

Hence, the SHG tensor for the transient state is

$$\chi(2\omega;\omega,\omega)_{\alpha\beta,\text{transient}} = \chi(2\omega;\omega,\omega)_{\alpha\beta,\bar{4}} + \delta\chi(2\omega;\omega,\omega)_{\alpha\beta,2}$$

$$= \begin{bmatrix} 0 & 0 & 0 & \chi_{14}+\delta\chi_{14} & \chi_{15}+\delta\chi_{15} & 0 \\ 0 & 0 & 0 & -\chi_{15}+\delta\chi_{24} & \chi_{14}+\delta\chi_{25} & 0 \\ \chi_{31}+\delta\chi_{31} & -\chi_{31}+\delta\chi_{32} & \delta\chi_{33} & 0 & 0 & \chi_{36}+\delta\chi_{36} \end{bmatrix} \quad (S2.4)$$

where $\delta$ indicates the transient contribution.

In addition, the coherent excitation of the E-symmetry mode $Q_E$ along the crystal $a$ axis results in a hyper-Raman scattering contribution to the time-resolved SHG. The susceptibility tensor for this process is written as

$$h_{\alpha\beta} = \begin{bmatrix} h_{11} & h_{12} & h_{13} & 0 & 0 & h_{16} \\ h_{16} & h_{22} & h_{23} & 0 & 0 & h_{12} \\ 0 & 0 & 0 & h_{23} & h_{13} & 0 \end{bmatrix} \quad (S2.5)$$

with the corresponding nonlinear polarization $P_i(2\omega) \propto h_{\alpha\beta} Q_E(\Omega) E_j(\omega) E_k(\omega)$. Here, $h_{\alpha\beta}$ is the Voigt form of $h_{ijk} = \frac{\partial}{\partial Q_E}\chi_{ijk}(2\omega;\omega,\omega)$, and $Q_E(\Omega)$ is the infrared-active phonon coordinate with frequency $\Omega$.

The nonlinear polarization, including the hyper-Raman response, can be expressed by expanding the second-order nonlinear susceptibility $\chi^{(2)}$ in the phonon coordinate $Q_E(z,t)$

$$P(z,t) = \epsilon_0 \chi^{(2)} E^2(z,t) = \epsilon_0 \left( \chi^{(2)}|_{Q_{E0}=0} + \frac{\partial \chi^{(2)}}{\partial Q_{E0}}\bigg|_{Q_{E0}=0} Q_E(z,t) \right) E^2(z,t) \quad (S2.6)$$

where $\epsilon_0$ presents the vacuum permittivity and $E(z,t)$ is the electric field of the fundamental beam. The rectified signal component at the detector becomes

$$I_{\text{rect}} \propto \left( \left|\chi_0^{(2)}\right|^2 + 2h^2 |Q_{E0}|^2 \right) \quad (S2.7)$$

where $\chi_0^{(2)} = \chi^{(2)}|_{Q_{E0}=0}$, $h = \frac{\partial \chi^{(2)}}{\partial Q_{E0}}\bigg|_{Q_{E0}=0}$, and $Q_{E0}$ denotes the amplitude of $Q_E(z,t) = Q_{E0} e^{i(k_{Q_E} z - \Omega_E t)} + c.c.$ with the phonon wave vector of $k_{Q_E}$. This expression includes approximations $\omega_{HR\pm} \sim \omega_2$, $n_{HR\pm} \sim n_2$, and $\Delta k_{HR\pm} \sim \Delta k$, where $\omega_2 = 2\omega_1$ corresponds to the second harmonic of the fundamental frequency $\omega_1$, and $\omega_{HR\pm} = 2\omega_1 \pm \Omega_E$ represent the hyper-Raman sidebands associated with the driven E phonon mode of frequency $\Omega_E$. $n_i = n(\omega_i)$ denote the refractive index, $\Delta k = 2k_1 - k_2$, and $\Delta k_{HR\pm} = \pm k_{Q_E} + 2k_1 - k_{HR\pm}$ where $k_i$ are wavevectors. We derived Eq. S2.7 following the approach outlined in Reference 2.

For a probe polarization oriented at an angle $\theta$ with respect to the $c$ axis, the coefficients can be substituted by the tensor elements in Equations S2.4 and S2.5 as

$$\chi_0^{(2)} \rightarrow (\chi_{31} + \delta\chi_{31}) \sin^2\theta + \delta\chi_{33} \cos^2\theta \quad (S2.8)$$

$$h \rightarrow h_{13} \sin\theta \cos\theta$$

Assuming $\delta\chi_{31}$ to be negligible, the rectified intensity reduces to

$$I_{\text{rect}} \propto \delta\chi_{33}^2 \cos^4\theta + 2(\chi_{31}\delta\chi_{33} + h_{13}^2 |Q_{E0}|^2) \sin^2\theta \cos^2\theta \quad (S2.9)$$

The first term originates from $\delta\chi_{33}$ and produces two horizontal lobes in the SHG polarimetry. The second term contains the homodyne contribution of the hyper-Raman response and the heterodyne mixing between SHG electric fields originating from the static coefficient $\chi_{31}$ and the transient component $\delta\chi_{33}$.

### S2.3. Fluence dependence of the rectified SHG components

To rationalize the fluence dependencies shown in Fig. 3b of the main text, we consider the rectified SHG intensity as given in Eq. S2.9.

For the probe polarization $\theta = 0°$, the expression reduces to $I_{\text{rect}} \propto \delta d_{33}^2$. The transient coefficient $\delta\chi_{33}$ scales with the symmetry breaking induced by rectification of the coupled mode $Q_B$, hence $\delta\chi_{33} \propto Q_B$. Because $Q_B$ is driven by $Q_E^2$ and $Q_E$ scales linearly with the pump electric field $E_{\text{pump}}$, one obtains $I_{\text{rect}} \propto \delta\chi_{33}^2 \propto E_{\text{pump}}^4 \propto F^2$ where $F$ presents the excitation fluence.

For the probe polarization $\theta = 45°$, where $\cos^4\theta = \sin^4\theta = 0.25$, only the second term in Eq. S2.9 needs to be considered. As $\chi_{31}$ is a static coefficient, the second term scales with the excitation as $\delta\chi_{33} \propto E_{\text{pump}}^2$. Moreover, the hyper-Raman response of the resonantly excited mode ($h_{13}^2 Q_{E0}^2$) is expected to exceed the contribution from the coupled mode ($\delta\chi_{33}^2$), and the static coefficient $\chi_{31}$ is anticipated to be larger than the transient component $\delta\chi_{33}$. Consequently, the second term dominates, leading to an overall dependence $I_{\text{rect}} \propto E_{\text{pump}}^2 \propto F$.

### S2.4. Fitting procedures

### S2.4.1. Fits to the excitation fluence dependent data

Figure 3b of the main text shows the fluence dependent rectified SHG intensity components for probe polarizations of $\theta = 0°$ and $45°$.

At 45°, the time traces were fitted with an error-function–exponential model.

$$I_{rect}(t) = A \exp\left(-\frac{t-t_0}{\tau}\right)\frac{1}{2}\left[1 + \text{erf}\left\{2\sqrt{2}\left(\frac{t-t_0}{t_r}\right)\right\}\right] + C \qquad (S2.10)$$

where $A$ is the amplitude, $t$ the pump-probe time delay, $t_0$ the temporal overlap (time zero), $\tau$ the 1/e decay constant, erf the error function, $t_r$ the rise time, and $C$ a constant.

Figure S2.2a shows a representative trace at the maximum excitation fluence, together with the corresponding fit. The extracted amplitude $A$ is plotted as a function of excitation fluence in Fig. S2.2b and follows $I_{\text{rect}} \propto F$, consistent with the prediction presented in Section S2.3.

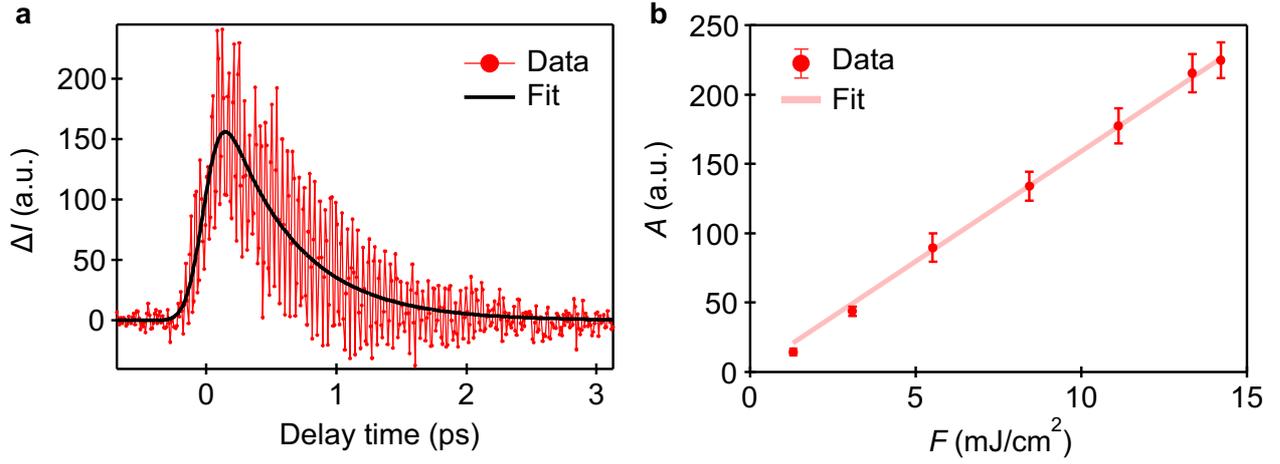

**Fig. S2.2. Fluence dependence at the probe polarization 45°. a**, Time trace at the maximum excitation fluence. **b**, Extracted amplitude (*A*) as a function of the excitation fluence *F*.

At 0°, the time traces were fitted with a double error-function–exponential model to capture two distinct signal contributions

$$I_{rect}(t) = A_f \exp\left(-\frac{t-t_f}{\tau_f}\right)\frac{1}{2}\left[1 + \text{erf}\left\{2\sqrt{2}\left(\frac{t-t_f}{t_f}\right)\right\}\right]$$
$$+ A_s \exp\left(-\frac{t-t_s}{\tau_s}\right)\frac{1}{2}\left[1 + \text{erf}\left\{2\sqrt{2}\left(\frac{t-t_s}{t_s}\right)\right\}\right] + C \quad (S2.11)$$

where *A* is the amplitude, *t* the pump-probe delay time, $t_0$ the temporal overlap (time zero), $\tau$ the 1/e decay constant, erf the error function, $t_r$ the rise time, and *C* the constant. The subscripts *f* and *s* denote the fast and slow decay components, respectively.

Figure S2.3a shows a representative time trace and the corresponding fit at the maximum excitation fluence. The amplitudes of the fast and slow decay components are plotted as a function of excitation fluence in Figs. S2.3b and c, respectively. The fast decay component (see Fig. S2.3b) follows $I_{\text{rect}} \propto F^2$, as discussed in Section S2.3, and is attributed to the transient reduction of crystal symmetry by rectification of the B-symmetry modes. In contrast, the slow component (see Fig. S2.3c) scales as $I_{\text{rect}} \propto F$, which is likely due to leakage of the 45° probe polarization channel into this signal component.

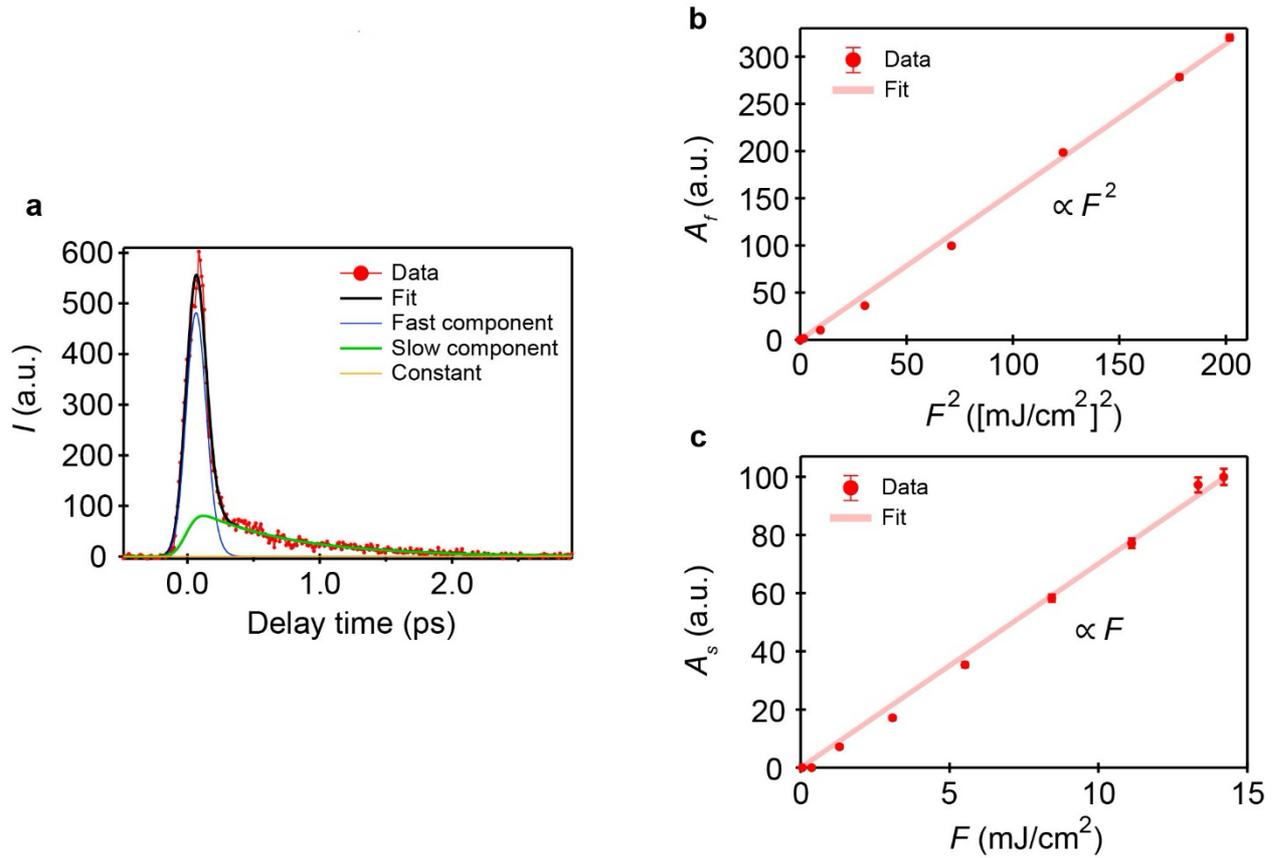

**Fig. S2.3. Fluence dependence at the probe polarization 0°. a**, Time trace at the maximum excitation fluence. **b**, Amplitude $A_f$ of the fast component as a function of $F^2$ where $F$ presents the excitation fluence. **c**, Amplitude $A_s$ of the slow component as a function of $F$.

**S2.4.2. Fits to the time-resolved SHG polarimetry data**

Figure 3a of the main text shows the rectified components of the photo-induced changes in the SHG polarimetry. Figure S2.4 shows the complete data set underlying this analysis for the example of the maximum excitation fluence.

At each probe polarization, the time-dependent SHG intensity change was fitted using Eq. S2.10, and the probe polarization dependent amplitudes $A$, extracted from these fits, were used in the polar plots shown in the main text. Then, Eq. S2.9 was used to fit the polarimetry pattern. The same procedure was applied to the two polarimetry data sets acquired at lower excitation fluences.

Strictly, in the vicinity of a probe polarization of 0°, a double error-function–exponential function would be more appropriate to account for the two distinct signal contributions (see discussion in Section S2.4.1). However, for consistency we employed the single error-function–exponential model across all the polarization angles.

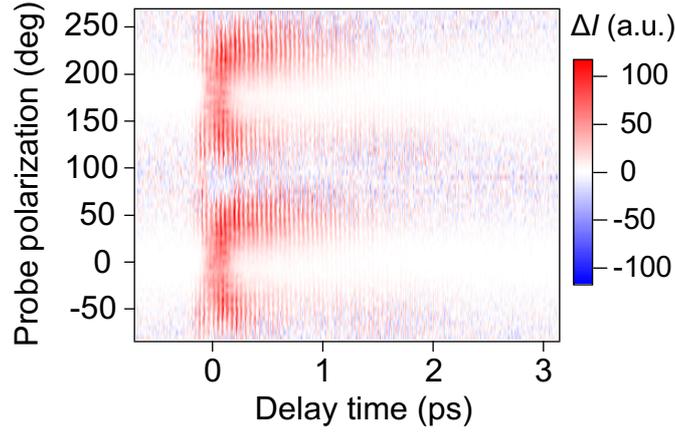

**Fig. S2.4. Time-delay and probe-polarization-angle dependent changes in the SHG intensity at the maximum excitation fluence.**

### S3. Theoretical modeling

#### S3.1. Ab-initio modeling (density functional theory (DFT))

We performed first-principles calculations in the framework of density functional theory (DFT) to explore the phonon excitation spectrum, the anharmonic lattice coupling coefficients, and the optical response of $BPO_4$ within the following technical and numerical settings. We used the Vienna Ab-initio Simulation Package (VASP) 6.5 DFT implementation[3–5] and the Phonopy software package[6] for the phonon calculations.

Within VASP, our computations utilized pseudopotentials generated within the Projector Augmented Wave (PAW) method[7]. Specifically, we took the configured default pseudopotentials for B $2s^22p^1$, P $3s^23p^3$, and O $2s^22p^4$ potentials and applied the Generalized Gradient Approximation for solids (PBEsol[8]) for the exchange-correlation potential. After convergence testing, we adopted a 7x7x5 Monkhorst-Pack[9] k-point mesh for Brillouin-zone sampling and a plane-wave energy cutoff of 600 eV for the structural relaxation and phonon calculations. Self-consistent calculations were iterated until the total-energy change was below $10^{-8}$ eV. All phonon-mode and anharmonic-coupling-constant calculations were performed using a $2 \times 2 \times 2$ conventional supercell of $BPO_4$.

The starting point of our first-principles investigation was the tetragonal unit cell of $BPO_4$. We first determined the DFT equilibrium volume of this cell to be $V_{cell} = 131.7$ Å$^3$, with lattice parameters $a = b = 4.43$ Å and $c = 6.71$ Å. The equilibrium atomic positions were found at the following Wyckoff sites: B at 2d, P at 2a, and O at 8g with internal coordinate (0.132,0.258,0.127). This structure was then used as the basis for all subsequent calculations.

#### S3.2. Calculation of piezoelectric coefficient

To determine the magnitude of the piezoelectric coefficient and to obtain the structural changes induced by stress, we performed two sets of calculations. First, we manually applied stress by varying the *a*-lattice constant while relaxing all remaining structural degrees of freedom, namely the *b*- and *c*-axis lattice parameters as well as the internal atomic coordinates. These calculations were repeated for a series of *a*-lattice constants, with each structure relaxed using the same numerical settings as for the equilibrium state.

From these calculations, we extracted both the stress along the *a* axis and the electrical polarization induced by the structural distortions, which allowed us to directly determine the piezoelectric coefficient $d_{31}$ = 0.13 (μC/(cm$^2$·GPa)). As a cross-check, we also used the internal routines of VASP to compute the piezoelectric stress tensor and the compliance tensor. From these quantities, we obtained the full piezoelectric strain tensor, which has the following form:

$$d_{ij} = \begin{bmatrix} 0 & 0 & 0 & 0.47 & -0.31 & 0 \\ 0 & 0 & 0 & 0.31 & 0.47 & 0 \\ -0.15 & 0.15 & 0 & 0 & 0 & 1.24 \end{bmatrix} \left(\frac{\mu C}{cm^2 \cdot GPa}\right) \quad \text{(S3.1)}$$

Here, we utilized the DFT compliance tensor

$$C_{ij} = \begin{bmatrix} 0.85 & 0.31 & -0.32 & 0 & 0 & 0 \\ 0.31 & 0.85 & -0.32 & 0 & 0 & 0 \\ -0.32 & -0.32 & 0.79 & 0 & 0 & 0 \\ 0 & 0 & 0 & 1.78 & 0 & 0 \\ 0 & 0 & 0 & 0 & 1.56 & 0 \\ 0 & 0 & 0 & 0 & 0 & 1.56 \end{bmatrix} \left(\frac{\%}{GPa}\right) \quad \text{(S3.2)}$$

Assuming a fracture-limited maximum elastic strain of 1% in BPO$_4$, this corresponds to a stress of approximately 1.2 GPa, as estimated from $\epsilon_1 = C_{11}\sigma_{11}$. Here, $\sigma_1$ denotes the applied stress and $\epsilon_1$ the resulting strain.

Lastly, the internal atomic distortions for the applied pressure on the *a* axis are listed in Table S3.1.

**Table S3.1.** The positions and displacements of atoms in the conventional BPO$_4$ unit cell, upon the application of stress on the a-axis.

| Atom | x (pm) | y (pm) | z (pm) | $\delta x$ (pm/GPa) | $\delta y$ (pm/GPa) | $\delta z$ (pm/GPa) |
|------|--------|--------|--------|---------------------|---------------------|---------------------|
| B1   | 0      | 221.51 | 167.68 | 0.00                | 0.00                | -0.30               |
| B2   | 221.51 | 0      | 503.04 | 0.00                | 0.00                | -0.30               |
| P1   | 0      | 0      | 0      | 0.00                | 0.00                | 0.45                |
| P2   | 221.51 | 221.51 | 335.36 | 0.00                | 0.00                | 0.45                |
| O1   | 58.48  | 114.54 | 85.08  | 2.24                | 0.04                | -0.02               |
| O2   | 384.53 | 328.48 | 85.08  | -2.24               | -0.04               | -0.02               |
| O3   | 114.54 | 384.53 | 585.64 | 0.14                | -2.07               | -0.05               |
| O4   | 328.48 | 58.48  | 585.64 | -0.14               | 0.10                | -0.05               |
| O5   | 279.99 | 336.04 | 420.44 | 2.24                | 0.04                | -0.02               |
| O6   | 163.03 | 106.97 | 420.44 | -2.24               | -0.04               | -0.02               |
| O7   | 336.04 | 163.03 | 250.28 | 0.14                | -2.07               | -0.05               |
| O8   | 106.97 | 279.99 | 250.28 | -0.14               | 2.07                | -0.05               |

**S3.3. Comparison of anharmonic couplings to B modes between phono-electric and piezoelectric effects**

For the equilibrium structure, we computed the phonon eigenmodes, frequencies and mode effective charges (see Table S3.2). Then, we calculated the anharmonic coupling constants between the resonantly driven $E$-mode and all the four $B$-modes. This was done following the approach of our previous work[1], in which we freeze in combinations of the two modes in question and fit the resulting energy landscape using a polynomial expression. The resulting coefficients, which are also used in our dynamical simulations, are listed in Table S3.2. Here, $\beta_{Bi}$ is the coefficient for the nonlinear potential of $V = -\beta_{Bi}Q_{Bi}Q_E^2$ where $Q_{Bi}$ and $Q_E$ are phonon coordinates for modes B$_i$ and E, respectively. $\beta_{Bi}$ is identical to $\alpha_i^{(8)}$ in Eqs. S3.7 and S4.1.

**Table S3.2. Parameter values for the $E$ and $B$ phonon modes obtained from first-principles calculations.**

|  | $f_i$ (THz) | $Z^*$ (e/u$^{1/2}$) | $\beta_{Bi}$ (meV/(u u$^{1/2}$ Å)$^3$) | $\zeta_j$ |
|---|---|---|---|---|
| E mode | 26.8 | 1.17 | | |
| B$_1$ mode | 15.4 | 0.87 | -6.6 | 51.5 |
| B$_2$ mode | 17.0 | 0.39 | 16.6 | -75.7 |
| B$_3$ mode | 26.6 | 1.08 | 897.6 | -126.6 |
| B$_4$ mode | 30.4 | 1.46 | 373.9 | 83.6 |

Finally, we map the eigendisplacement of the pressure-induced distortion listed in Table S3.1 onto the phonon eigenmodes. We do so by computing

$$\zeta_i = \frac{1}{|Q_p|}\sum_j [\hat{Q}_{Bi}]_j \cdot [Q_p]_j \quad (S3.3)$$

where $\hat{Q}_{B_i}$ is the phonon eigenvector of the $i$-th $B$ mode, $Q_p$ is the displacement pattern from Table S3.1, the index $j$ runs over all atoms, and |·| denotes the norm of the vectors. The resulting coefficients are also listed in Table S3.2. Figure S3.1a shows the B-mode response, $Q_{Bi} = \zeta_i\hat{Q}_{Bi}$, as a function of applied stress. Figure S3.1b presents the corresponding induced electrical polarizations obtained by weighting each mode $Q_{Bi}$ with its mode effective charge $Z^*_{Bi}$ (Table S3.2). Noteworthy, the sum of the induced polarizations by the four B modes in Fig. S3.1b nearly vanishes, resulting in a weak piezoelectric response. In contrast, in the case of nonlinear phononics, i.e. the phonoelectric effect, a different combination of B mode displacements yields a substantial net electrical polarization, as shown in Fig. S3.2. This behavior is also reflected in the signs of $\beta_{Bi}$ and $\zeta_i$ in Table S3.2: $\beta_{Bi}$ exhibits one negative and three positive contributions, whereas $\zeta_i$ comprises two negative and two positive terms.

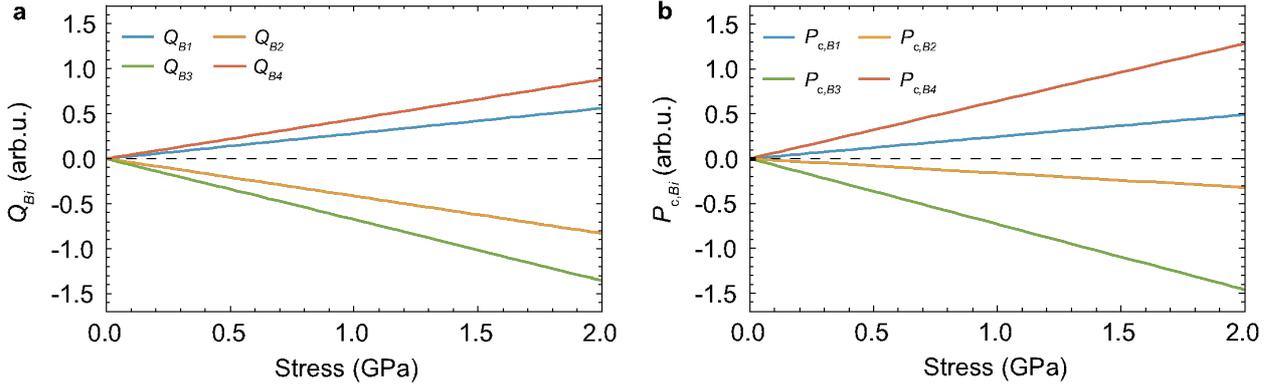

**Fig. S3.1. Mode-resolved lattice response and induced electrical polarization in the piezoelectric effect. a**, B mode response, $Q_{Bi} = \zeta_i \hat{Q}_{Bi}$, as a function of applied stress. **b**, Corresponding induced electrical polarization.

### S3.4. Calculation of electrical polarizations

To calculate the induced electric polarization through nonlinear phononics, we numerically solved the following equations of motion for the driven E mode ($Q_E$) and coupled B mode ($Q_B$) phonon coordinates.

$$\ddot{Q}_E = -2\gamma_E \dot{Q}_E - \frac{\partial V}{\partial Q_E} \tag{S3.4}$$

$$\ddot{Q}_{Bi} = -2\gamma_{Bi} \dot{Q}_{Bi} - \frac{\partial V}{\partial Q_{Bi}} \tag{S3.5}$$

Here, the subscript $i$ labels the four B modes, and $\gamma_E$, $\gamma_{Bi}$ denote phenomenological damping constants. The induced polarization can be evaluated by the equation

$$P_c = \sum_i Z^*_{B_i} Q_{B_i} \tag{S3.6}$$

where $P_c$ denotes the induced polarization along the c axis and $Z^*_{B_i}$ is the mode effective charge of the mode $B_i$.

We utilized the following potential term including higher order anharmonic terms as follows

$$V = (-1) \times \begin{bmatrix} \frac{1}{2}\omega_E^2 Q_{Ea}^2 + \alpha^{(1)} Q_{Ea}^4 + \alpha^{(2)} Q_{Ea}^6 + \alpha^{(3)} Q_{Ea}^3 Q_{Eb} + \alpha^{(4)} Q_{Ea}^2 Q_{Eb}^2 + \alpha^{(5)} Q_{Ea} Q_{Eb}^3 \\ + \sum_i \left( \frac{1}{2}\omega_{Bi}^2 Q_{Bi}^2 + \alpha_i^{(6)} Q_{Bi}^4 + \alpha_i^{(7)} Q_{Bi}^6 \right) \\ + \sum_i \left( \alpha_i^{(8)} Q_{Bi} + \alpha_i^{(9)} Q_{Bi}^2 \right) Q_{Ea}^2 + \sum_i \alpha_i^{(10)} Q_{Bi} Q_{Ea} Q_{Eb} + Z^*_{Ea} E_{pump} Q_{Ea} \end{bmatrix} \tag{S3.7}$$

where the coefficients $\alpha^{(j)}$ with $j = 1 - 10$ were obtained from DFT calculations. Here, $\omega_E$ and $\omega_{Bi}$ denote the natural frequencies of E and $B_i$ modes, $E_{pump}$ the electric field of the mid-infrared excitation pulse, $Z^*_{Ea}$ the mode effective charge of the E mode along a axis, and $Q_{Ea}$ and $Q_{Eb}$ the E mode phonon coordinate along a and b axis, respectively. Note that E modes along a and b axis are degenerate.

The parameter values for $\omega_i(=2\pi f_i)$, $Z_i^*$, $\gamma_i$, $\alpha_i^{(8)}(=\beta_{Bi})$ are from Table S3.2 and S3.3.

**Table S3.3. Parameter values of the damping coefficient $\gamma$ for E and B phonon modes, used in the calculations**

|  | $\gamma$ |
|---|---|
| E mode | $0.034 \times f_E$ |
| B$_1$ mode | $0.1 \times f_{B1}$ |
| B$_2$ mode | $0.1 \times f_{B2}$ |
| B$_3$ mode | $0.1 \times f_{B3}$ |
| B$_4$ mode | $0.1 \times f_{B4}$ |

The factor 0.034 in $\gamma_E$ was taken from the ratio $\gamma_E/\omega_{TO,E}$ obtained from fits to Fourier-transform infrared reflectivity (FTIR) measurements reported in Table S1 of Reference 1.

The pump electric field, $E_{pump}$, was modeled as a sinusoidal carrier with a frequency of 26.79 THz and a Gaussian envelope with a full width at half maximum (FWHM) pulse duration of $\sqrt{2} \times 165$ fs.

An example calculation of the polarization induced by the rectified B modes is shown in Fig. S3.2, for a peak electric field inside BPO$_4$ of 2 MV/cm with a pulse duration of $\sqrt{2} \times 165$ fs. These values correspond to fluences inside BPO$_4$ and in air of 0.93 mJ/cm² and 1.63 mJ/cm², respectively. The Fresnel transmission coefficient is $2/(1+\sqrt{\epsilon_\infty}) \approx 0.76$, where $\epsilon_\infty = 2.7$ in BPO$_4$[1] and the pulse duration is assumed to remain unchanged upon transmission from air to the crystal. Time traces of the B mode coordinates $Q_{Bi}(t)$ and the corresponding polarizations were smoothed using a Gaussian filter with FWHM of 51.8 fs.

To determine the excitation fluence dependence of the induced polarization, we varied the pump electric field while keeping the pulse duration fixed. The maximum induced electrical polarization obtained from the calculations was fitted with a linear function of the excitation fluence, as shown in Fig. S3.3. The fit yields

$$P_c = 0.051 \times F \tag{S3.8}$$

where $P_c$ is the induced polarization in μC/cm² and $F$ is the excitation fluence in mJ/cm². Note that the horizontal axes of Figs. 4 and S3.3 use the incident fluence in air.

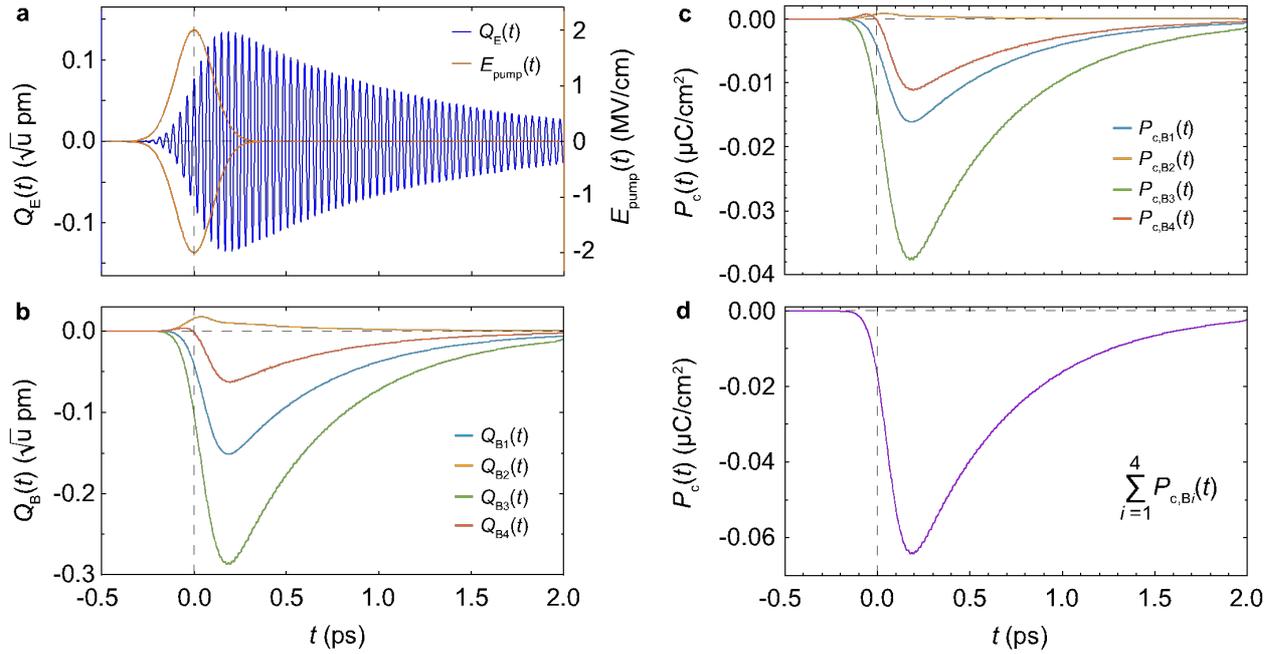

**Fig. S3.2. Example calculation of the electrical polarization induced by rectified B modes. a,** Envelope of the pump electric field and the time trace of the phonon coordinate of the excited E mode. Here, u denotes the atomic mass unit ($\approx 1.66 \times 10^{-27}$ kg). **b,** Time traces of the phonon coordinates of the coupled B modes. **c,** Time traces of the polarization along the $c$ axis of $BPO_4$ crystal arising from each B mode. **d,** Time trace of the total polarization obtained by summing all the polarizations shown in panel **c**.

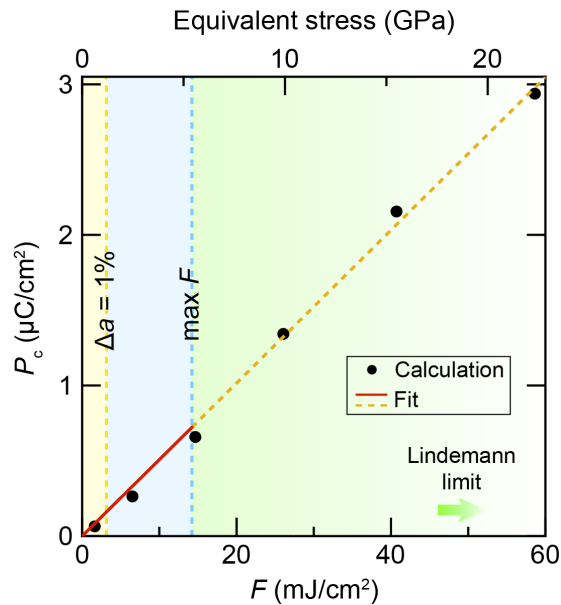

**Fig. S3.3. Comparison of the phonoelectric and piezoelectric effects.** Polarization values correspond to the maximum absolute value of the time trace (as in Fig. S3.2d) and were calculated at five different excitation fluences by varying the electric field while keeping the pulse duration fixed. Calculated points are shown as black dots, and a linear fit of the polarization versus the excitation fluence is overlaid. Vertical lines indicate the equivalent stress of 1.2 GPa (corresponding to 1 % strain, yellow) and the maximum excitation fluence used in the experiments (blue).

### S3.5. Calculation of the excitation electric field at the Lindemann stability criterion

We estimated the maximum permissible excitation field strength from a theoretical perspective. For this purpose, we analyze the eigenvector with the largest amplitude in the driven state, namely the directly excited $E$ mode. This mode predominantly modulates the P–O bond, which has an equilibrium length of 1.55 Å. At a phonon amplitude of 4 $u^{1/2}$ Å, the bond is elongated by 10%, which we take as the upper limit of the physically allowed amplitude according to the Lindemann stability criterion. For the excitation configuration described in Sec. S3.4, this phonon amplitude requires a peak electric field strength of 60 MV/cm, corresponding to the incident fluence of 1.47 J/cm². Using Eq. S3.8, the induced electrical polarization is estimated to be about 74 uC/cm².

## S4. Propagation speeds: Induced polarization vs. acoustic phonons

### S4.1. Calculation of the propagation speed of the induced polarization

To calculate the propagation speed of the induced electrical polarization, we performed numerical simulations based on the finite element method in COMSOL. Within a one-dimensional geometry, we solved Maxwell's equation together with Eqs. S3.4 – S3.6. In these simulations, we adopted the following potential of the form

$$V = (-1) \times \left[ \frac{1}{2}\omega_E^2 Q_E^2 + \sum_i \frac{1}{2}\omega_{Bi}^2 Q_{Bi}^2 + \sum_i \alpha_i^{(8)} Q_{Bi} Q_{Ea}^2 + Z_E^* E_{pump} Q_E \right] \quad (S4.1)$$

All parameter values were taken from Sections S3.3 and S3.4. The pump electric field is introduced at 1 ps and propagates through 400 μm of vacuum before entering the sample. Figures S4.1a,b show the spatiotemporal evolution of the E mode phonon-polariton and the induced polarization, respectively. The propagation speed of the induced polarization was extracted from the rise edge, as illustrated in Fig. S4.1c. A linear fit yields a speed of 41 μm/ps.

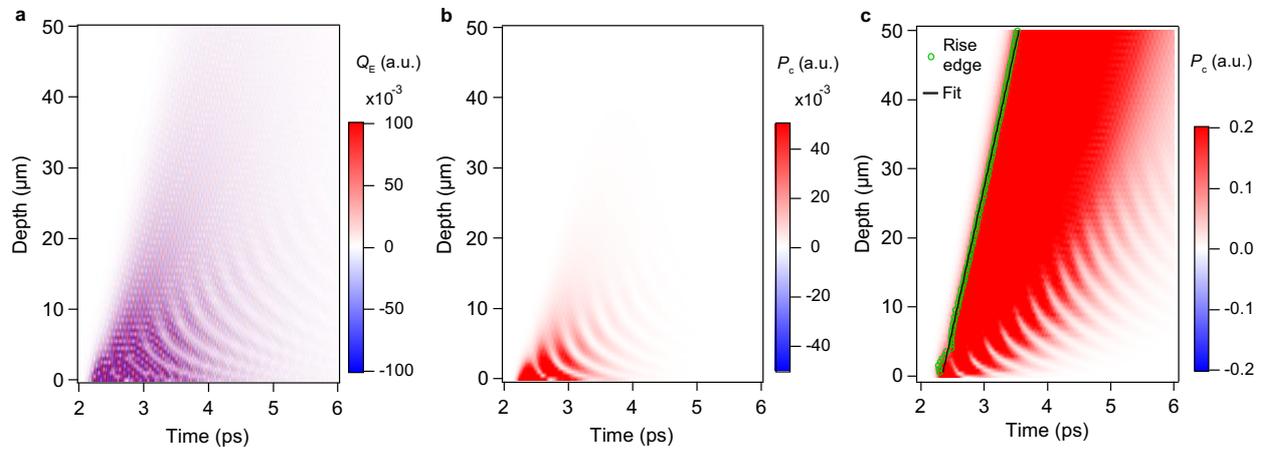

**Fig. S4.1. Propagation of the driven E mode phonon-polariton and the induced electrical polarization. a,b,** Spatiotemporal evolution of the E mode phonon-polariton $Q_E$ and the induced polarization $P_C$. **c,** Same data as in b, however the time traces at each depth are normalized independently to highlight the propagation speed of the nonlinear electrical polarization. The green solid line is a fit to the rising edge of the polarization.

### S4.2. Calculation of the propagation speed of the acoustic phonon

To estimate the propagation speed of the electrical polarization induced by piezoelectric effect, we evaluated the longitudinal acoustic phonon velocity $v_A$ in BPO$_4$ along the a axis, assuming an applied stress along the same direction. This is given by

$$v_A = \sqrt{\frac{S_{11}}{\rho}} \qquad (S4.2)$$

where $S_{11}$ (= 147.2 GPa) is the stiffness tensor element and $\rho$ (=2.80$^{10}$ g/cm³) is the mass density. The stiffness tensor $S_{ij}$ is obtained as the inverse of the compliance tensor given in Eq. S3.2,

$$S_{ij} = C_{ij}^{-1} = \begin{bmatrix} 147.2 & -38.0 & 43.4 & 0 & 0 & 0 \\ -38.0 & 147.2 & 43.4 & 0 & 0 & 0 \\ 43.4 & 43.4 & 160.5 & 0 & 0 & 0 \\ 0 & 0 & 0 & 56.2 & 0 & 0 \\ 0 & 0 & 0 & 0 & 64.0 & 0 \\ 0 & 0 & 0 & 0 & 0 & 64.0 \end{bmatrix} \text{(GPa)} \qquad (S4.3)$$

The resulting value of $v_A$ is 7.25 nm/ps.

## S5. Extinction depth of the mid-infrared excitation

For the discussion of the energy efficiency in the main text, the input energy density of the phono-electric effect ($w_{phono}$) is defined as $F/\delta$, where $F$ denotes the excitation fluence and $\delta$ the extinction depth of the mid-infrared pulses.

The frequency-dependent extinction depth $\delta(\nu)$ is given by

$$\delta(f) = \frac{1}{4\pi} \frac{\lambda_0(\nu)}{\kappa(\nu)} \qquad (S5.1)$$

where $\lambda_0(\nu)$ is the wavelength in vacuum and $\kappa(\nu)$ is the imaginary part of the dielectric function, obtained from fits to Fourier-transform infrared reflectivity (FTIR) measurements reported in Table S1 of Reference 1. The resulting $\delta(\nu)$ was weighted by the excitation spectrum $S_p(\nu)$ (shown above in Fig. S1.2) according to

$$\delta = \frac{\int \delta(\nu) S_p(\nu) d\nu}{\int S_p(\nu) d\nu} \qquad (S5.2)$$

The integration was performed over the frequency range of 20 THz to 40 THz. The resulting extinction depth of 1.4 um was used for the energy efficiency calculation.

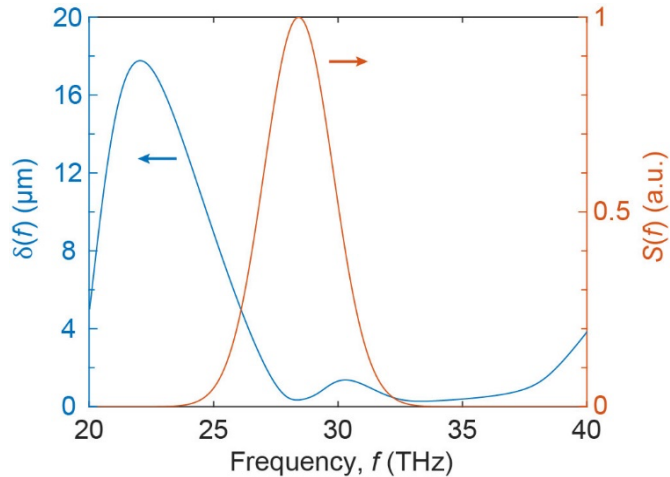

**Fig. S5.1. Frequency-dependent extinction depth $\delta(\nu)$ of the mid-infrared pulses, together with the excitation spectrum, $S_p(\nu)$.**

**References Supplementary Information**